# Atomic-scale studies of $Fe_3O_4$(001) and $TiO_2$(110) surfaces following immersion in $CO_2$-acidified water

Dr. Francesca Mirabella[1], Dr. Jan Balajka[1,2], Dr. Jiri Pavelec[1], Markus Göbel[1], Florian Kraushofer[1], Prof. Dr. Michael Schmid[1], Prof. Dr. Gareth S. Parkinson[1], and Prof. Dr. Ulrike Diebold[1*]

[1]*Institut für Angewandte Physik, Technische Universität Wien, A-1040 Wien, Austria*
[2]*Department of Chemistry and Chemical Biology, Cornell University, Ithaca, NY 14853 USA*

*corresponding author: diebold@iap.tuwien.ac.at

**Abstract**

Difficulties associated with the integration of liquids into a UHV environment make surface-science style studies of mineral dissolution particularly challenging. Recently, we developed a novel experimental setup for the UHV-compatible dosing of ultrapure liquid water, and studied its interaction with $TiO_2$ and $Fe_3O_4$ surfaces. Here, we describe a simple approach to vary the pH through the partial pressure of $CO_2$ ($p_{CO_2}$) in the surrounding vacuum chamber, and use this to study how these surfaces react to an acidic solution. The $TiO_2$(110) surface is unaffected by the acidic solution, except for a small amount of carbonaceous contamination. The $Fe_3O_4$(001)-($\sqrt{2}$ x $\sqrt{2}$)R45° surface begins to dissolve at a pH 4.0-3.9 ($p_{CO_2}$ = 0.8–1 bar) and, although it is significantly roughened, the atomic-scale structure of the $Fe_3O_4$(001) surface layer remains visible in scanning tunneling microscopy (STM) images. X-ray photoelectron spectroscopy (XPS) reveals that the surface is chemically reduced, and contains a significant accumulation of bicarbonate ($HCO_3^-$) species. These observations are consistent with Fe(II) being extracted by bicarbonate ions, leading to dissolved iron bicarbonate complexes ($Fe(HCO_3)_2$), which precipitate onto the surface when the water evaporates.

1. **Introduction**

Most of the existing theories of Earth-abundant-mineral degradation are based on kinetic measurements. For example, great progress towards a microscopic understanding of dissolution has been made using X-ray reflectivity to measure the dissolution rate in real time [1], and by high-speed frequency modulation AFM for direct imaging of the surface step edges during dissolution [2]. However, a mechanistic understanding of the dissolution process remains rudimentary. One reason is that atomic-scale studies require an initial state in which the surface structure and composition are precisely defined. Gaining such control of the system is usually achieved using of single crystals, and many surface analysis techniques require an ultrahigh vacuum (UHV) environment. Liquid water does not exist at room temperature in UHV, and immediately evaporates and freezes if inserted into vacuum. Integrating liquid water into UHV-based experiments thus requires the sample to be exposed to some ambient pressure environment prior to immersion [3,4,5,6]. This can cause contamination of the sample, which can lead to erroneous interpretation of experimental data [7]. Recently, we developed a method to integrate liquid water exposure with a surface science approach [8]. Essentially, UHV-prepared samples are exposed to ultrapure liquid water in a custom-designed cell, without the need for other venting gases (e.g. air, Ar, $N_2$). Before introducing the sample to the cell, water vapour is frozen on a cold finger located above the sample holder, and after sample transfer under UHV conditions, the ice melts and falls onto the UHV-prepared surface. We have shown in our prior work that metal oxide surfaces exposed to liquid water in this way exhibit minimal contamination when returned to UHV, and that $TiO_2(110)$ and $Fe_2O_3(012)$ are largely unaffected by pure water [7,9]. $Fe_3O_4(001)$, on the other hand, responds to liquid water exposure forming a new surface phase, which we interpreted as an Fe oxy-hydroxide layer based on scanning tunnelling microscopy (STM) and X-ray photoelectron spectroscopy (XPS) results [10]. In this paper, we vary the pH of the water drop in a controlled way through the dissolution of $CO_2$ in the water drop during immersion. Specifically, we show that the $Fe_3O_4(001)$ surface changes with decreasing pH, and begins to dissolve at pH ≈ 4. On the other hand, the $TiO_2(110)$ surface remains unaffected by an acidic solution, as expected on the basis of the Ti Pourbaix diagram [11].

Besides serving as a proof of principle, our results are relevant for the study of chemical weathering of Fe oxides, which plays a substantial role in the regulation of the atmospheric partial pressure of carbon dioxide ($p_{CO_2}$) [12,13,14,15,16,17]. The Fe Pourbaix diagram [18] predicts solvated Fe(II) to be the thermodynamically favored species at ambient conditions (zero applied potential and pH=7), but the omnipresence of Fe(III) and Fe(II, III) oxides in nature suggests that the dissolution process is kinetically limited [19]. Kinetic studies have indeed shown magnetite dissolution to be facet-specific, with dissolution rates increasing in the sequence {110} < {100} < {111} [20]. One reason is that surface $Fe^{2+}$ ions are thought to be necessary for the dissolution reaction to proceed to completion [21], while Zinder et al. [16] proposed that the degradation process is kinetically controlled by the detachment of an Fe center from the surface and is enhanced in the presence of complex-forming organic ligands. Our results show similar behavior to that reported in ref. [16], and highlight the synergistic effect of protons and counter-ions concentration in solution on the dissolution rate.

## 2. Experimental Details

The experiments were performed on a natural $Fe_3O_4$(001) single crystal (SurfaceNet GmbH), prepared in UHV by cycles of 1 keV $Ar^+$ sputtering and 900 K annealing. Every other annealing cycle was performed in an $O_2$ environment ($p_{O_2}$ = 5×10$^{-7}$ mbar, 20 min) to maintain the stoichiometry of the crystal selvedge. The surface analysis was performed in a UHV system with a base pressure <10$^{-10}$ mbar, furnished with a commercial Omicron SPECTALEED rear-view optics and an Omicron UHV STM-1. XPS data were acquired using non-monochromated Mg Kα x-rays and a SPECS PHOIBOS 100 electron analyser at grazing emission (70° from the surface normal). All STM images presented in this work were corrected for distortion and creep of the piezo scanner, as described in ref. [22].

The magnetite surface was exposed to an ultra-pure liquid-water drop in a custom-designed side chamber described in detail elsewhere [8]. A small cryostat ("cold finger") is used for preparing the water drop. Milli-Q water (Milipore, 18.2 MΩ cm) is initially purified by freeze-pump-thaw cycles, and introduced as vapour through a valve into the evacuated side chamber. The vapour pressure of the water in the reservoir was adjusted at 6 mbar by stabilizing the temperature of the bath slightly above 0 °C to avoid condensation on the walls of the side chamber. To avoid replacing weakly bound adsorbates at the walls of the side chamber by adsorption of $H_2O$, experiments with a sample were only conducting after "washing" the walls by three cycles of filling the side chamber with 6 mbar water vapor at RT. In the actual experiment, the water freezes on the cold finger, which is held at cryogenic temperature and acts as a localized cold spot. Following the evacuation of the side chamber using a $LN_2$-cooled cryo-sorption pump, the UHV-prepared sample is transferred directly below the cold finger. After closing the valve between the side chamber and the rest of the UHV system, the cold finger is heated until the icicle thaws, and ultra-pure water drops onto the sample surface. The chamber is then re-evacuated, and the sample transferred back to the main chamber for analysis. This procedure ensures that the surface is exposed to water only, and no air contamination occurs.

Acidic water drops were prepared by additionally introducing $CO_2$ (purity 99.995% from Linde HiQ® MINICAN) into the side chamber while the water drop was already on the sample surface. The pressure of the $CO_2$ was monitored and controlled by a capacitance pressure gauge. The specified $CO_2$ pressure was kept for ca. 20 min to allow for the gas to dissolve into the water. We assume that we reached equilibrium in this time-scale because longer exposure did not lead to different results. The evacuating process, as well as the transfer back to the UHV system, were identical to that described for the pH-neutral water drop.

Details about the equilibria involved in the dissolution of carbon dioxide in water and how this affects the pH of the acidic solution are in the supplemental information (SI). Briefly, when $CO_2$ dissolves in water, it exists in chemical equilibrium with carbonic acid ($H_2CO_3$). The composition of a carbonic acid solution is fully determined by the $p_{CO_2}$ above the solution. The amount of dissolved $CO_2$ increases with increasing $p_{CO_2}$ surrounding the water drop, and it dominates by 2 orders of magnitude over the next highest component, $HCO_3^-$. The concentration of dissolved $CO_3^{2-}$ is always negligible in the range of $p_{CO_2}$ considered here.

## 3. Results

### 3.1 Fe$_3$O$_4$(001) - Scanning Tunneling Microscopy and Low Energy Electron Diffraction

The UHV-prepared $Fe_3O_4$(001) surface exhibits a (√2 × √2)R45° reconstruction due to an interstitial tetrahedrally coordinated iron in the second layer ($Fe_{tet}$), which replaces two octahedrally coordinated iron atoms ($Fe_{oct}$) in the third layer [23]. This reconstruction, also known as the subsurface cation vacancy (SCV) structure, is the most stable termination for the $Fe_3O_4$(001) surface over a wide range of $O_2$ chemical potentials relevant for UHV studies [23]. Figure 1(a) shows an STM image of the as-prepared clean $Fe_3O_4$(001) surface, which exhibits the characteristic rows of protrusions running in the [110] direction due to the surface Fe cations. Bright protrusions on the Fe rows are caused by surface OH groups (i.e. hydrogen atoms adsorbed at surface oxygen atoms), which modify the density of states of the nearby Fe cations, causing them to appear brighter in empty-states STM images [24,25]. Other common defects visible on the clean surface image are antiphase domain boundaries, which appear as meandering line defects, and defects that appear similar to two neighboring hydroxyl groups, but do not diffuse at room temperature. These are caused by an additional Fe atom being present in the subsurface layer, which again modifies the density of states of the surface atoms [24,26].

Figure 1(b) shows an STM image of the Fe$_3$O$_4$(001) surface following exposure to an ultra-pure water drop for 20 minutes. The surface exhibits bright chains with an apparent height of ≈2.1 Å covering ≈40% of the surface. We recently interpreted these chains as representative of an Fe-(oxy)hydroxide phase, formed when two or more O$_{water}$H groups from dissociated water coordinate tetrahedrally coordinated Fe$_{tet}$ atoms extracted from the subsurface [10]. The surface Fe rows along the <110> directions remain visible in the surrounding surface, but all have almost the same apparent height due to extensive hydroxylation of the surface oxygen lattice. The necessary H is the counterpart of the OH resulting from dissociative water adsorption. The corresponding LEED pattern exhibits a (1 × 1) symmetry (inset), which is known to occur when the H coverage reaches two H atoms per unit cell [24,27]. The complete hydrogenation of the surface caused by the dissociative water adsorption was invoked as the reason why the growth of the oxyhydroxide features terminates at approx. 40%: when no sites are available to accommodate the H atoms, no water dissociation is possible [10].

Figures 1(c)-(d) show the Fe$_3$O$_4$(001) surface imaged following exposure to acidic water drops performed with a $p_{CO_2}$ = 20 mbar (pH≈4.8) and $p_{CO_2}$ = 1 bar (pH≈3.9), respectively. The former surface (c) exhibits bright chains consistent with the growth of an iron(oxy)hydroxide phase, as observed at pH 7, but with the coverage of the bright chains decreased to ≈30% of the surface. In the corresponding LEED pattern (inset), the (√2 × √2)R45° reconstruction spots remain absent, and a (1 × 1) symmetry is observed. Following exposure to the more acidic solution (pH 3.9), the bright chains are completely absent and the surface appears rough in STM, suggesting that etching has occurred. Nevertheless, terraces with rows of protrusions rotated 90° with respect to each other (typical for monoatomic steps on magnetite surfaces) remain visible, suggesting the atomic-scale structure of the surface layer remained intact. LEED shows a (1 × 1) pattern of similar quality to that observed at neutral pH. Isolated protrusions with an apparent height between 2.3-3.2 Å are common on the surface, but occupy positions both on the rows and in between them. Based on the XPS data shown in the next section, we suspect these to be related to carbonaceous species.

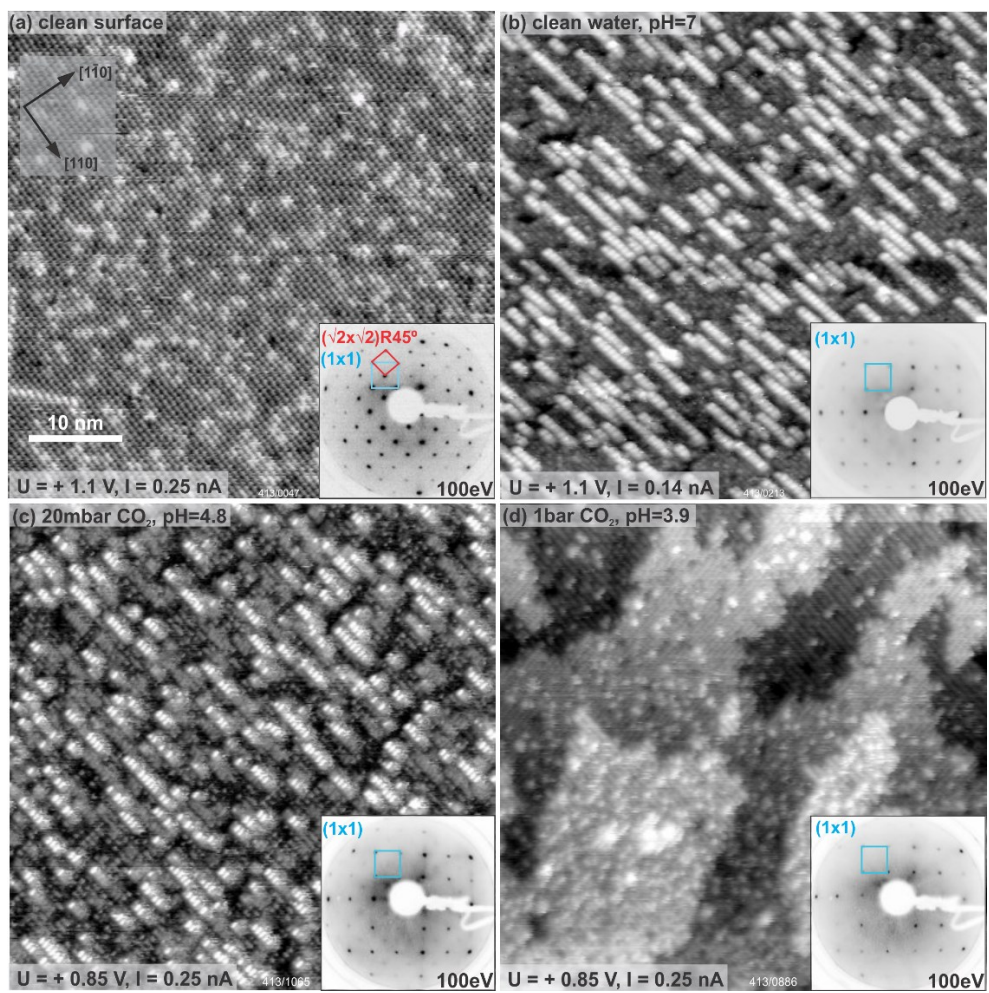

*Figure 1. Room-temperature STM images of the Fe$_3$O$_4$(001) surface before and after exposure to liquid water in varying background pressures of CO$_2$ (50 × 50 nm$^2$).* As prepared in UHV (a), after exposure to ultra-pure liquid water (b), after exposure to an acidic solution at (c) pH 4.8, and (d) pH 3.9. Corresponding LEED patterns are shown in the insets, with the reciprocal (√2 × √2)R45° and (1 × 1) unit cells drawn in red and blue, respectively.

### 3.2 $Fe_3O_4$(001) - X-ray-photoelectron Spectroscopy

The XPS spectra (Mg Kα, 70° grazing emission) in Figures 2(a)-(c) were acquired on the as-prepared $Fe_3O_4$(001) surface, and following exposure to water in different background pressures of $CO_2$. Data corresponding to the STM images in Fig. 1 are shown, as well as a spectrum for a $p_{CO_2}$ of 800 mbar (pH≈4.0). The corresponding STM image shows roughness similar to Fig. 1(d) (see Fig. S1). The regions of interest, namely Fe *2p*, C *1s*, and O *1s*, are shown in panels (a), (b) and (c) respectively.

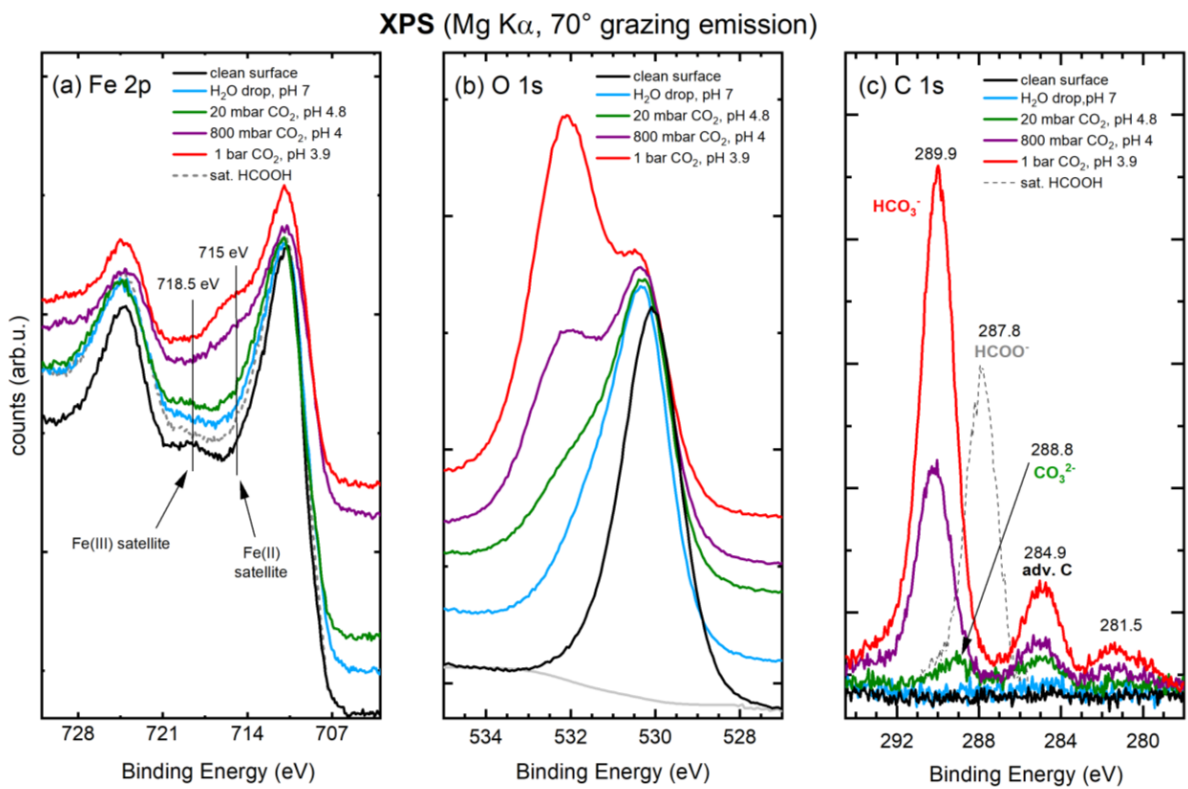

*Figure 2. XPS spectra (Mg Kα, 70° grazing emission).* Fe 2p (a), O 1s (b) and C 1s (c) regions before and after exposing the $Fe_3O_4$(001)-(√2 × √2)R45° surface to liquid water as well as to acidic solutions at pH 4.8, 4.0, and 3.9. Spectra after exposure to gas-phase HCOOH, resulting in a saturation coverage of bidentate formate[27] are shown for comparison (grey, dashed). Spectra in (a), (b) are shifted vertically for clarity.

The Fe *2p* spectrum of the SCV reconstructed $Fe_3O_4$(001) surface is enriched in $Fe^{3+}$ with respect to the bulk [23], which is most evident in the appearance of a strong $Fe^{3+}$ shake-up satellite at 718.5 eV [29]. Following exposure to ultrapure water ($p_{CO_2}$ = 0, light blue curve), the overall intensity of the Fe 2p peak area is slightly attenuated and the $Fe^{3+}$-related satellite peak at 718.5 eV disappears, while the $Fe^{2+}$-related intensity at 715 eV increases. The enhancement of the $Fe^{2+}$ in the surface region is also visible as an increased $Fe^{2+}$ shoulder on the low-binding-energy side (at 708.1 eV) of the main $2p_{3/2}$ peak [29]. The reduction of the surface Fe becomes more pronounced following exposure to progressively more acidic solutions, with a systematic growth of the $Fe^{2+}$-related [29] satellite at 715 eV as well as a subtle increase of the shoulder at 708.1 eV. Figure 2(a) also show the Fe *2p* signal following exposure to saturation coverage of HCOOH (grey dashed line in Fig. 2(a)). Formic acid dissociates on the $Fe_3O_4$(001) surface at RT to form bidentate formate species and hydroxyl groups [27], causing surface reduction. This is visible in the Fe *2p* peak, which, similarly to the case of water adsorption, shows both the missing Fe(III) satellite at 718.5 eV and a slight increase of the Fe(II) shake-up feature at 708.1 eV.

The O *1s* peak from the clean $Fe_3O_4$(001) surface is centred at 530.1 eV, and is slightly asymmetric due to the metallic nature of the oxide [30]. After the interaction with ultrapure liquid water, the O *1s* spectrum shows an additional component at 531.6 eV (a fit for this peak is reported in ref. [10]), which is assigned to hydroxyl groups due to dissociative water adsorption [10]. Molecular water would be expected at 533 eV, but it is not observed. As the pH of the water drop decreases to 4.8, the surface oxygen signal decreases in intensity and a new contribution appears at 532.2 eV, which is in the region typical for C-O bonds [31]. The latter feature grows as the pH decreases to 4.0 and becomes higher in intensity than the surface O signal when the pH of the water drop decreases to 3.9. A fit for the O 1s peak measured after exposing the surface to the acidic solutions at pH 4.8, 4, and 3.9 is showed in Fig. S2.

The C *1s* spectrum of the UHV-prepared surface is free from carbon within our detection limit (< 0.05 C atoms per (√2 × √2)R45° unit cell). The water-exposed surface shows a small peak at 284.9 eV corresponding to adventitious carbon. Comparing the latter peak with a reference spectrum of saturation coverage of formate adsorbed on the $Fe_3O_4$(001) surface (grey dashed line in Fig.2(c)), which has a known density of two C atoms per (√2 × √2)R45° unit cell [27], we estimate an adventitious carbon coverage of 0.1 C atoms per (√2 × √2)R45° unit cell ($1.4 \times 10^{13}$ C atoms/cm$^2$). Following the interaction with progressively more acidic solutions, the surface exhibits higher adventitious carbon signals, with the highest C density of 0.6 C atoms per (√2 × √2)R45° unit cell ($8.4 \times 10^{13}$ C atoms/cm$^2$) after exposure to acidic solutions at pH 3.9. More interestingly, the surface exposed to the acidic water drop at pH 4.8 exhibits a new broad peak at 288.8 eV. This is clearly shifted from the formic acid reference, and was previously assigned to surface carbonate species [31,32,33,34]. As the pH is decreased to 4.0 and to 3.9, a peak at 289.9 eV appears, indicating the presence of a surface bicarbonate ($HCO_3^-$) species [31,32,33,35]. In addition, we observe a small feature at lower binding energy (281.5 eV) which is 8.4 eV apart from the main bicarbonate peak. This corresponds to the spacing between the $\alpha_{1,2}$ and the $\alpha_3$ lines of the Mg anode, and therefore the peak can be attributed to an x-ray satellite of the $\alpha_{1,2}$ signal at 289.9 eV.

In Figure 3, we plot the integrated intensity of the XPS peak at 289.9 eV (red points) against the $p_{CO_2}$ surrounding the water drop during the experiment. With no applied $CO_2$, the peak is completely absent, and it increases slowly going from 20 mbar, to 200 and 800 mbar. Then, a steep increase is observed when going from 800 to 1000 mbar. This corresponds in STM to a significant roughening of the surface morphology, which we interpret as the result of etching. On the opposite y-axis we plot the calculated concentration of $HCO_3^-$ in the solution, assuming that the standard equilibrium equations are directly applicable. Clearly, both curves grow with increasing $p_{CO_2}$ dissolved into the water drop; however, the surface bicarbonate concentration increases significantly when the $p_{CO_2}$ goes from 800 mbar to 1000 mbar, deviating from the square-root trend followed by the [$HCO_3^-$] in solution. The reaction that leads to the formation of the surface bicarbonate (which correlates with the surface dissolution) is clearly not linearly dependent on the concentrations of reactants ([$HCO_3^-$] in solution), so we can speculate that the surface-bicarbonate concentration measured in XPS as a function of $p_{CO_2}$ might be an indication of the reaction rate.

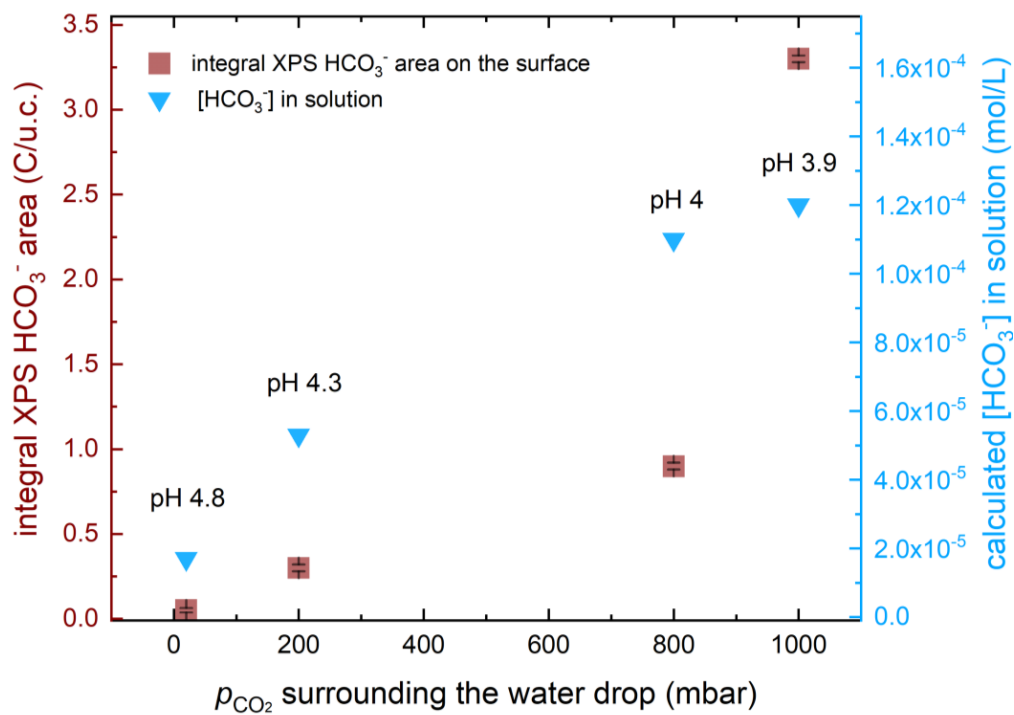

*Figure 3. Comparison of the $HCO_3^-$ species concentrations in solution (light blue) and on the surface after solution evacuation (red).* The bicarbonate concentration in solution was calculated according to the equilibria described in eq. (3)-(4) in the SI. The bicarbonate concentration on the surface and error bars were obtained by calculating the integral peak area, as well as the corresponding standard deviation, of the related signal measured in XPS using the software CasaXPS.

To test whether the acidic solution is required to form the stable bicarbonate surface species, we performed a control experiment. The as-prepared sample was exposed to 800 mbar $CO_2$ gas without the liquid water drop (Fig. 4). No bicarbonate is detected in this experiment, but a relatively large adventitious C peak is observed at 284.5 eV. This suggests that the presence of the water drop protects the surface from contamination by impurities in the $CO_2$ gas in the other experiments. A small signal at 288.9 eV is assigned to carbonate species, likely formed by the interaction of the $CO_2$ with surface defects [33,34,36]. Finally, we explored the thermal stability of the surface bicarbonate species created by the pH 4.0 exposure. Figure 5 shows that the bicarbonate-related peaks in the C 1s and O 1s XPS spectra decreased in intensity as the temperature increased up to 150 °C and completely disappeared following heating at 200 °C.

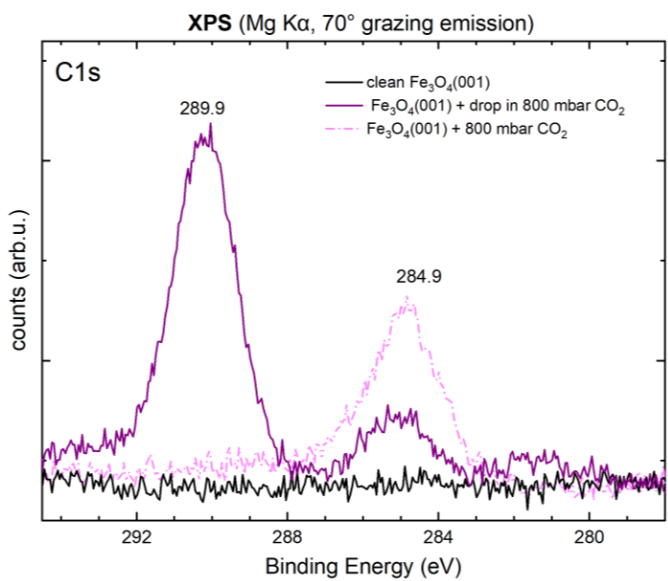

*Figure 4. Control experiment exposing $Fe_3O_4$(001) to $CO_2$ in the absence of water. XPS spectra (Mg Kα, 70° grazing emission) of the C 1s region of a clean $Fe_3O_4$(001) surface (black) and after exposure to an acidic solution at pH 4.0 (violet), and after exposure to 800 mbar $CO_2$ (pink, dashed). The solution at pH 4.0 was prepared by dissolving 800 mbar of $CO_2$ into a water drop.*

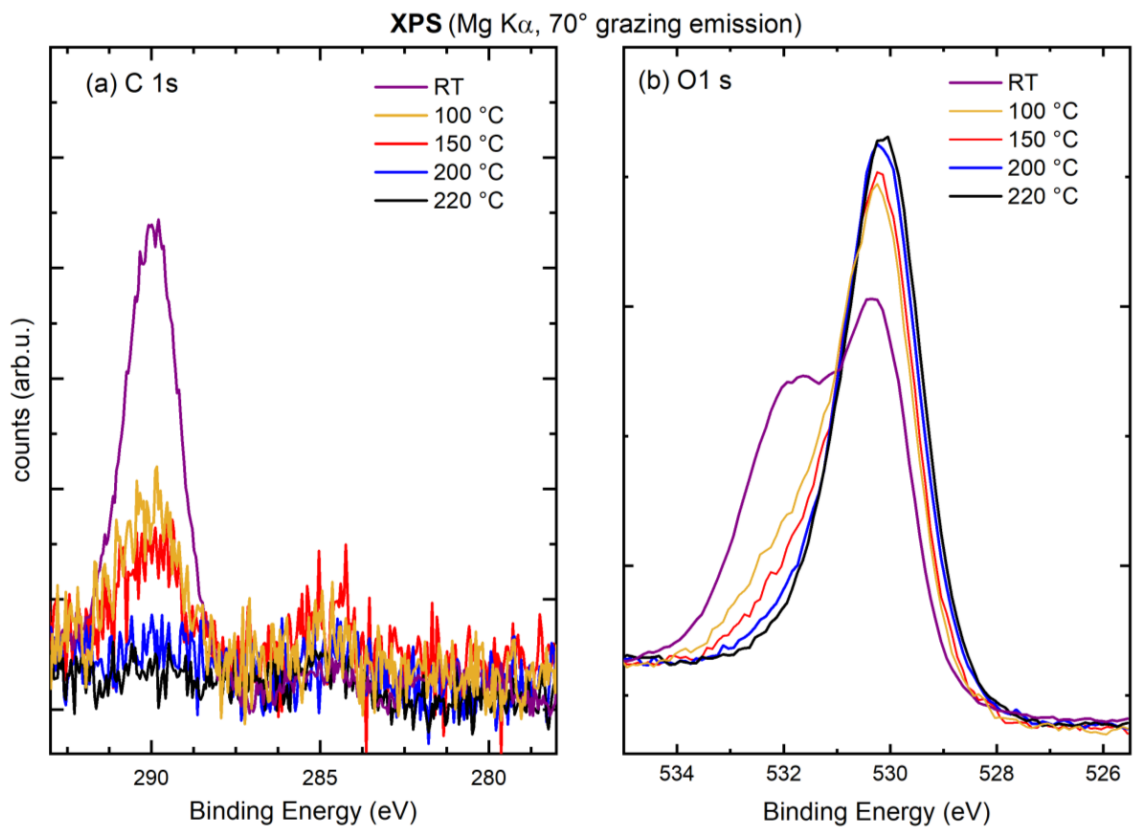

*Figure 5. Thermal stability of the carbonaceous species formed by exposing $Fe_3O_4$(001) to $CO_2$-acidified water.* XPS spectra (Mg Kα, 70° grazing emission) of the C 1s (a), O 1s (b) of a $Fe_3O_4$(001)-(√2 × √2)R45° surface after exposure to an acidic solution at pH 4.0 at room temperature (violet) and then heated up to 100 °C (yellow), 150 °C (red), 200 C (blue), 220 °C (black).

### 3.3 $TiO_2$(110) - STM and LEED

Figure 6 shows a set of STM images of the $TiO_2$(110) surface. Empty-states STM images of a UHV-prepared $TiO_2$(110) surface are dominated by five-fold coordinated surface titanium atoms, which appear as bright rows running along the [001] direction with a 6 Å spacing [37] (Fig. 6(a)). Figure 6(b) shows a STM image of the $TiO_2$(110) surface following exposure to an ultra-pure water drop for 20 minutes. The resulting surface looks indistinguishable to the clean one in STM, with the exception of a low density of contaminants, which are imaged as white aggregates (white blobs). This result is in agreement with Balajka et al. [7], who showed that liquid water does not modify the $TiO_2$(110) surface and retains its UHV-like (1 × 1) structure upon water exposure (see LEED in the inset). Figure 6(c) shows the $TiO_2$(110) surface imaged following exposure to an acidic water drop in which the $p_{CO_2}$ was set to 800 mbar (pH≈4). The resulting $TiO_2$(110) surface remains overall intact and continues to exhibit a (1 × 1) periodicity in LEED (inset). However, it is possible to distinguish in STM a fractional monolayer coverage (~0.15 ML) of adsorbate species, all located in equivalent sites and have an apparent height of 1.5-1.7 Å. These species appear similar in STM to the ones observed on the air-exposed surface reported in ref. [7], where it has been shown that air contamination leads to the adsorption of carboxylic acids (present in air in ppb concentrations) which arrange on the surface in a (2 × 1) structure. Perhaps coincidently, the coverage of these species is very similar to the saturation coverage of polarons expected for a reduced $TiO_2$(110) surface [38]. To identify the nature of the features observed in STM, we also performed XPS experiments on this surface.

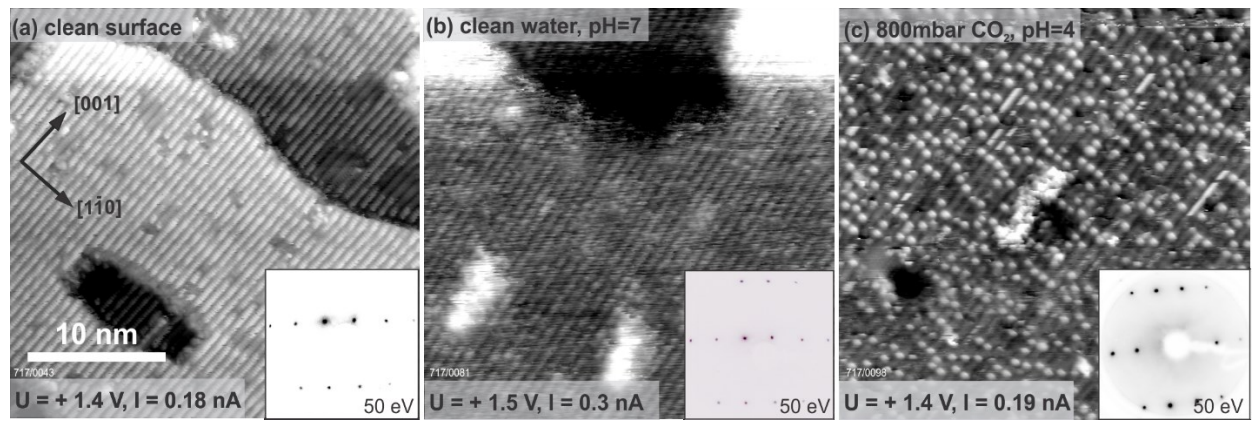

*Figure 6. Room-temperature STM images of the $TiO_2$(110) surface before and after exposure to pure liquid water and liquid water in background high pressure of $CO_2$. (a) as prepared in UHV, (b) after exposure to ultra-pure liquid water, (c) after exposure to the acidic solution at pH 4. Corresponding LEED patterns are shown in the insets.*

### 3.4 $TiO_2$(110) - X-ray-photoelectron Spectroscopy

In Figure 7 we show C1s XPS data for the surface imaged in Figure 6 (c). For comparison, the result of the same experiment performed on $Fe_3O_4$(001) is also shown (violet, dashed).

The C *1s* spectra of the UHV-prepared $TiO_2$(110) surface is free from carbon within the detection limit of the instrument. When the surface is exposed to an acidic solution at pH 4, the C *1s* region of the XP spectrum shows two small peaks at 289.3 eV and at 284.9 eV. The signal at 289.3 eV has been previously assigned to carboxylate species [7]. Based on a saturation coverage of 0.5 ML HCOOH on $TiO_2$ (one formate per two surface unit cell, not shown) [39], the C density for the 289.3 eV species and the adventitious C corresponds to 0.035 and 0.05 C per surface unit cell ($1.7 \times 10^{13}$ and $2.4 \times 10^{13}$ C atoms/cm$^2$) respectively.

As in the case of magnetite, we performed a control experiment to study the effect of high-pressure $CO_2$ on the $TiO_2$ surface without a water drop. The surface was exposed to 800 mbar $CO_2$ gas and the XP signal of the C 1s region is reported in Figure 7, green. The C1s signal shows a peak at 289.3 eV of comparable intensity to the one obtained following acidic solution interaction, and an adventitious carbon peak at 284.9 eV, corresponding to a higher C density of 0.08 C/surface unit cell ($3.8 \times 10^{13}$ C atoms/cm$^2$). These results suggests that the acidic environment is not responsible for the formation of the adsorbate species and would speak in favor of carboxylates formation, in agreement with Balajka et al. [7]

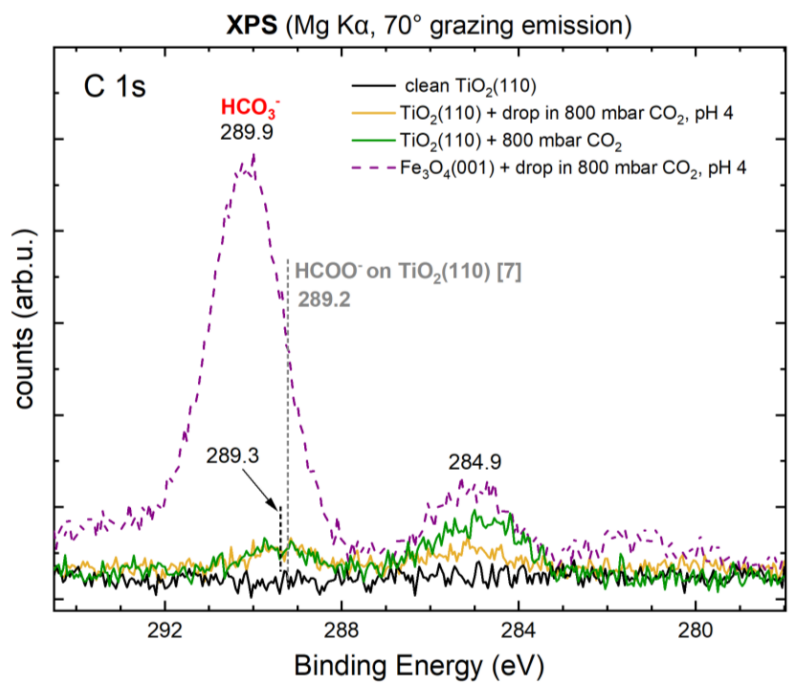

*Figure 7. TiO$_2$(110) in comparison to Fe$_3$O$_4$(001).* *XPS spectra (Mg Kα, 70° grazing emission) of the C 1s region of a clean TiO$_2$(110) surface (black) and after exposure to an acidic solution at pH 4.0 (yellow), and after exposure to 800 mbar CO$_2$ (green). For comparison, the C 1s signal of the Fe$_3$O$_4$(001) after exposure to an acidic solution at pH 4.0 (violet, dashed) is also shown. The solution at pH 4.0 was prepared by dissolving 800 mbar of CO$_2$ into a water drop.*

## 4. Discussion

Our results suggest that it is possible to change the pH of an ultra-pure water drop in a controlled way by varying the $p_{CO_2}$ in the surrounding environment. As case studies, we have characterized the atomic-scale structure of the Fe$_3$O$_4$(001) and TiO$_2$(110) surface following immersion to acidic water drops. Both surfaces exhibit very little adventitious carbon contamination when returned to UHV, and the experiments maintain a similar degree of cleanliness despite the presence of high pressures of CO$_2$. Control experiments performed using CO$_2$ gas only (no water drop) show significantly more adventitious carbon contamination, which suggests the water drop acts as a protective film shielding the surface from hydrocarbon contaminants in the gas. Interestingly, the presence of the water drop does not change the amount of carboxylate formed on $TiO_2$, suggesting that these species are probably a small amount of carbonaceous contamination.

Generally speaking, our observations are consistent with the Fe and Ti Pourbaix diagrams [11,18]: dissolved $Fe^{2+}$ is the most stable form of Fe in acidic conditions, meaning dissolution should be expected. $TiO_2$, on the other hand, is thermodynamically stable in the pH range considered here, so this surface should remain intact, as observed. While we do not measure the pH directly; the fact that we observe the onset of significant roughening of the $Fe_3O_4$(001) surface at pH≤4 is consistent with prior studies of Fe oxide dissolution rates [17], and this gives us confidence that the system reaches equilibrium in the course of the experiment. Since the behavior of the two metal oxides are different and consistent with earlier literature, we conclude that the results of our experiment can be interpreted in terms of the interaction of the surfaces with the acidic water drop.

As mentioned above, the bright chains of protrusions observed after exposing $Fe_3O_4$(001) to water at pH 7 have been previously interpreted as an iron-(oxy)hydroxide phase, which forms due to the dissociative adsorption of water [10]. The protrusions contain $O_{water}H$ groups coordinated to Fe atoms from the surface, and the growth is limited by the number of reactive O sites, which are required to accommodate the accompanying proton. As the pH is lowered from pH 7, it seems likely that an increasing proportion of the reactive O sites are protonated directly from solution, which would explain why the coverage of the bright $O_{water}H$ related protrusions decreases, until they are completely absent at pH = 3.9. From STM it appears that the surface remains fully protonated (the Fe rows remain clearly visible), but it is difficult to confirm this from XPS because the OH O *1s* signal overlaps with bicarbonate, and the $Fe^{2+}$ signal becomes much too large to be attributed solely due to protonation.

To explain the origin of the surface bicarbonate, and the additional $Fe^{2+}$, we propose the following scenario: In Figure 1(d), we see that the surface exposed to water at pH 3.9 has a significantly higher step density than after exposing to pH 4.8, which suggests that the surface dissolution rate increased significantly at this condition, and that Fe extraction does not happen uniformly. More likely, Fe is preferentially removed along defects sites [17,21,40]. The observation of Fe(II) and $HCO_3^-$ in XPS (Figs. 2(a-c)) can result from the formation of iron-bicarbonate species, involving the extraction of a Fe from the surface which then goes into solution as $Fe(HCO_3)_2$. When the drop evaporates, the $Fe(HCO_3)_2$ would precipitate out and land on the surface where it accumulates.

If the bicarbonate were simply adsorbed from the solution in a monolayer coverage we would expect the bicarbonate peak in the C 1s region (289.9 eV) to have a C density of 2 C atoms per (√2 × √2)R45° unit cell [27]. Calibrating against the saturation coverage of formate, which likely binds in a similar fashion, we find that the bicarbonate signal shown in Fig. 2(c) corresponds to 3.3 C atoms per (√2 × √2)R45° unit cell. Turning to the Fe *2p* region, we note that the $Fe^{2+}$ contribution of the pH 3.9 surface is significantly larger than the one obtained at pH 7, where the surface was already completely hydroxylated. Moreover, the $Fe^{2+}$ signal is much larger than obtained for the $HCOO^-$ monolayer reference (Fig. 2(a)). Given the similarity of formate and bicarbonate, it is hard to imagine that an adsorbed layer is responsible for such extensive reduction. Previous kinetic studies by Zinder et al.[16] found that the dissolution of iron oxides proceeds faster in the presence of complex-forming organic ligands. Following their arguments, we propose that both $H^+$ and the $HCO_3^-$ ions in solution at pH 3.9 play a similar role as organic ligands in the dissolution. First, the adsorption of protons would polarize the neighbour Fe-O bond [41,28] and reduce surface $Fe^{3+}$ to $Fe^{2+}$. The now available Fe(II) is complexed by the bicarbonate ions, which act as ligands and form iron bicarbonate surface species (Fe(HCO₃)₂) [33,35,41], according to the equation $Fe^{2+} + 2HCO_3^- \rightarrow$ Fe$(HCO_3)_2$. The formation of these species results in the extraction of the Fe from the surface into the solution. Upon evacuation of the liquid, the iron bicarbonate species ultimately precipitate out of solution and accumulate onto the surface, where they are detected in XPS. The re-deposition of Fe-containing species is probably not uniform (areas where the liquid evaporates last may get a higher coverage), thus the surface structure observed by STM after etching and evaporation of the drop (Fig. 1d) is not necessarily representative for all of the surface.

Finally, one constraint of our current setup is that the use of high pressures for any gas is limited to 1000 mbar, which in the case of CO₂ limits the pH to ≈3.9. It would be interesting to see if the bicarbonate concentration decreased again at lower pH, as the maximum in the dissolution rate for iron oxides has been shown to depend on the counter anion [17]. For example, the maximum in the dissolution rate of hematite in citric acid occurs in the range of pH 4-5, whereas in oxalic acid it occurs at pH 1-2 [42]. For goethite in presence of oxalates, the optimum pH is in the range of pH 2-4 [43].

In Fig, 3, we see a significant increase in the surface bicarbonate signal when the $p_{CO_2}$ dissolved into the water drop is in the range 800-1000 mbar, suggesting that the proton concentration (pH) is not the only factor at play.

**Conclusions**

Our study demonstrates that it is possible to tune the pH of an ultra-pure water drop by varying the partial pressure of $CO_2$ in the surrounding environment, while keeping the high purity of a UHV-based surface science experiment. We have used this method to explore how two metal oxide surfaces, $Fe_3O_4$(001) and $TiO_2$(110), react to immersion in acidic water at the atomic scale.

Our experimental observations are consistent with the earlier predictions of the Fe and Ti Pourbaix diagrams. $TiO_2$ is predicted to be thermodynamically stable in the pH range considered in this work, and indeed we observe that the surface remains intact following exposure to an acidic solution at pH 4. On the other hand, the $Fe_3O_4$(001) surface changes with reducing pH and in presence of $HCO_3^-$ ions, and begins to dissolve at pH ≈ 4. Following earlier kinetic models, we propose that the magnetite dissolution process occurs involving three consecutive reactions, namely, surface protonation, ligand adsorption, and consequent metal detachment into solution. The chemical environment provided by the dissolution of $CO_2$ into the water drop favours the formation of protons and bicarbonates ($HCO_3^-$) in solution. At pH 4-3.9, the surface protonation and consequent formation of surface iron-bicarbonate species weakens the surface Fe-O bonds and favours the detachment of a Fe centre into solution. After evacuation of the liquid, the $Fe(HCO_3)_2$ precipitate from the solution and accumulate onto the surface, where are detected in XPS. Our approach opens the door to atomic-scale studies of degradation at the solid-liquid interface and can, in principle, be applied to any surface.

**Acknowledgements**

This work was supported by the European Union under the A-LEAF project (732840-A-LEAF), by the Austrian Science Fund FWF (Project 'Wittgenstein Prize, Z250-N27), and by the European Research Council (ERC) under the European Union's HORIZON2020 Research and Innovation program (ERC Grant Agreement No. [864628].

**Keywords**

$CO_2$; dissolution; magnetite; $TiO_2$; water drop